# Self-organization mechanism in Bridgman-grown MnBi$_2$Te$_4$/(Bi$_2$Te$_3$)$_n$: influence on layer sequence and magnetic properties

Paweł Skupiński,*[a] Kamil Sobczak,[b] Katarzyna Gas,[a,d] Anna Reszka,[a] Yadhu K. Edathumkandy,[a] Jakub Majewski,[c] Krzysztof Grasza,[a] Maciej Sawicki,[a,e] and Agnieszka Wołoś.[c]

a. Institute of Physics, Polish Academy of Sciences, Aleja Lotników 32/46, PL-02668 Warsaw, Poland.
b. Faculty of Chemistry, Biological and Chemical Research Centre, University of Warsaw, ul. Zwirki i Wigury 101, 02-089 Warsaw, Poland
c. Faculty of Physics, University of Warsaw, Pasteura 5, 02-093 Warsaw, Poland.
d. Center for Science and Innovation in Spintronics, Tohoku University, 2-1-1 Katahira, Aoba-ku, Sendai 980-8577, Japan.
e. Laboratory for Nanoelectronics and Spintronics, Research Institute of Electrical Communication, Tohoku University, 2-1-1 Katahira, Aoba-ku, Sendai 980-8577, Japan

* Author to whom any correspondence should be addressed.
E-mail: pawel.skupinski@ifpan.edu.pl

The growth of high-quality magnetic topological insulator crystals by the Bridgman method remains challenging due to thermodynamic limitations inherent to this technique. Nevertheless, this approach continues to provide bulk materials with significantly reduced free carrier concentrations compared to epitaxial methods. Here, we investigate the Inverted Vertical Bridgman growth of MnBi$_2$Te$_4$/(Bi$_2$Te$_3$)$_n$ crystals, with particular emphasis on the structural ordering of MnBi$_2$Te$_4$ septuple layers within the Bi$_2$Te$_3$ quintuple-layer matrix and its influence on magnetic properties. Through a detailed analysis of growth dynamics, we identify four distinct stages, including a turbulent flow regime promoting pure MnBi$_2$Te$_4$ phase, rapid MnTe precipitation reducing Mn content in the melt, a stationary growth phase supporting ordered stacking of septuple and quintuple layers, and a final stage marked by flow cessation and defect formation. We demonstrate that septuple layer spacing is inversely correlated with MnTe supersaturation due to the diffusion-limited incorporation in stationary growth phase. Magnetic characterization reveals antiferromagnetic ordering in pure MnBi$_2$Te$_4$ phase and in MnBi$_2$Te$_4$/Bi$_2$Te$_3$ heterostructure, with ferromagnetism emerging for wider septuple layer spacing. We determine critical temperatures for observed antiferromagnetic and ferromagnetic phase transitions and magnetic anisotropy constants for ferromagnetic samples. Our findings highlight key growth parameters governing magnetic and structural quality, offering a pathway to scalable synthesis of layered topological insulators with tunable magnetic properties.

## Introduction

Impurity segregation during Bridgman crystal growth is a well-known phenomenon, typically leading to a gradual enrichment or depletion of the melt in specific components, and resulting in spatial gradients in impurity-sensitive physical properties [1]. While commonly discussed in the context of charge carrier concentration, segregation effects involving magnetic dopants such as manganese (Mn) can also give rise to pronounced variations in magnetic behavior.

Mn-doped bismuth telluride (Bi$_2$Te$_3$) represents aparticularly intriguing system, where Mn atoms spontaneously self-organize within the Bi$_2$Te$_3$ matrix to form structurally and magnetically distinct septuple layers (SLs) of manganese bismuth telluride (MnBi$_2$Te$_4$). These SLs consist of a central Mn atomic plane exhibiting ferromagnetic alignment [2–5], and when embedded within the topological Bi$_2$Te$_3$ host, they give rise to magnetically and topologically nontrivial superlattices. Such heterostructures have emerged as a promising platform for realizing exotic quantum states, including the quantum anomalous Hall effect and axion insulator phases [2, 3, 6–10].

The host material, Bi$_2$Te$_3$, crystallizes in a layered rhombohedral structure composed of quintuple layers (QLs) stacked along the *c*-axis via weak van der Waals interactions. Upon Mn incorporation, partial substitution and local rearrangement produces a range of (MnBi$_2$Te$_4$)/(Bi$_2$Te$_3$)$_n$ heterostructures with varying stacking periodicity. The spacing, distribution, and integrity of these layers have been shown to critically influence the resulting magnetic and electronic properties [2–5].

While epitaxial approaches have been instrumental in probing the intrinsic physics of these materials, bulk crystals grown via the Bridgman technique offer the advantage of reduced free carrier concentrations and larger sample volumes. However, such crystal synthesis is inherently influenced by thermodynamic and kinetic factors, often resulting in stacking disorder, Mn segregation, and structural inhomogeneities that can mask or modify intrinsic material properties [5].

In this work, we focus on bulk MnBi$_2$Te$_4$/(Bi$_2$Te$_3$)$_n$ crystals grown using the Inverted Vertical Bridgman Method (IVBM) [11], which are composed solely of manganese, bismuth, and tellurium. Despite having the same overall chemical composition, different sections of the ingot exhibit markedly distinct magnetic and structural properties due to nanoscale variations in the stacking sequence and local manganese content. We investigate the mechanisms driving nanostructure formation under IVBM growth conditions and identify the key stages governing the self-organization of SLs and QLs. Our findings reveal the critical role of flow dynamics and melt composition in controlling SL spacing, stacking order, and resulting magnetic behavior, offering a pathway toward controlled synthesis of magnetically ordered topological materials.

## Experimental

Crystals were grown using the Inverted Vertical Bridgman Method from high-purity (99.999 %) Bi and Te from Alfa Aesar. and a synthesized in-house MnTe compound, with its characterization reported in [12]. Bi, Te, and MnTe were mixed in a weight ratio of 4.2:3.9:1.9. The convective mixing, initial synthesis and homogenization were conducted in vacuum-sealed quartz ampoules ($10^{-6}$ Torr) placed in a ten-zone vertical furnace for 24 hours with a temperature gradient ranging from 1270 K (bottom) to 1020 K (top). The melt was then stabilized at an constant temperature of 910 K along the entire length of the ampoules for 2 hours before initiating the final crystal growth. Ampoules were moved upward at 0.8 mm/h through a 15 K/cm gradient to a zone maintained at 570 K. Finally, the crystals were annealed at 575 K for 24 hours. The resulting ingots were typically 37 mm in length and 14 mm in diameter.

Structural characterization was carried out using two instruments: (i) An FEI Talos F200X Transmission Electron Microscope (TEM) operating at 200 kV, equipped with a High-Angle Annular Dark Field (HAADF) detector. The TEM was also fitted with an Energy-Dispersive X-ray (EDX) spectroscope (Super-X system) featuring four Silicon Drift Detectors (SDDs), enabling high-resolution chemical composition analysis. (ii) A Hitachi SU-70 Scanning Electron Microscope (SEM), equipped with a Thermo Scientific EDX spectrometer incorporating a SDD and the Noran System 7 for elemental analysis.

Magnetic resonance studies were conducted using a Bruker ELEXSYS-E580 spectrometer at X-band (9.5 GHz) with magnetic fields up to 1.6 T and temperatures down to 5 K, controlled by a continuous-flow Oxford helium cryostat. Using magnetic field modulation and lock-in detection, the resonance signal corresponds to the field derivative of absorbed microwave power. As electron paramagnetic, ferromagnetic, and antiferromagnetic resonances were measured, the results are collectively termed magnetic resonance.

Magnetic measurements were performed using a Quantum Design MPMS XL Superconducting Quantum Interference Device (SQUID) magnetometer equipped with a Reciprocating Sample Option (RSO) to enhance the signal-to-noise ratio. Measurements were carried out over the temperature range of 2–350 K. Small crystalline fragments (weighing a few milligrams) were mounted on 4 × 5 × 0.5 mm³ silicon substrate plates, using strongly diluted GE varnish to ensure mechanical stability. The samples were positioned such that the external magnetic field was applied along the crystallographic *c*-axis. All measurements were performed in accordance with established protocols for high-precision SQUID magnetometry, ensuring accurate detection of weak magnetic signals and minimizing experimental artifacts [13].

## Results and discussion

### Structural properties

Four samples were selected from an ingot grown by the IVBM, taken at distances of 34–36 mm (Sample A), 20–22 mm (Sample B), 10–13 mm (Sample C), and 0–3 mm (Sample D) from the bottom of the ampule. The corresponding TEM and EDX images are shown in Fig. 1.

Both the $Bi_2Te_3$ and its derivative $MnBi_2Te_4$ have layered crystalline structures, with the layers perpendicular to the crystal's *c*-axis (Fig. 1a). The $Bi_2Te_3$ QLs consist of five mono-atomic layers arranged along the *c*-axis in the following sequence: Te-Bi-Te-Bi-Te. For $MnBi_2Te_4$ SLs, the layer sequence is Te-Bi-Te-Mn-Te-Bi-Te. Atoms within QLs (or SLs) are bound by a mixture of covalent and ionic bonds while adjacent block layers interact by weak van der Waals forces, separated by van der Waals gaps [14–19]. TEM-EDX (Fig. 1b-f) reveals varying sequences of QLs and SLs along the crystal.

Sample A shows pure $MnBi_2Te_4$ phase (Fig. 1b) and locally alternating $MnBi_2Te_4$/$Bi_2Te_3$ inclusions (Fig. 1c). Sample B features SLs separated by one or two QLs (Fig. 1d) with dominating $MnBi_2Te_4$/$Bi_2Te_3$ phase. Further along the crystal, in Sample C, the SL spacing increases. A recurring pattern emerges: every three SLs tend to group with a spacing of (7 ± 2) QLs (Fig. 1e). Sample D, near the ingot end, exhibits increased structural disorder. SLs, spaced by 4–5 QLs, interchange with QLs via Mn-rich multilayers that intersect QLs at an angle (Fig. 1f). Notably, the spacing between SLs remains preserved.

Owing to the complexity of the superlattice structure, Mn concentration can be analyzed in three distinct regimes: within the QLs, the SLs, and as an average over the full superlattice. The average Mn concentration varies along the crystal, following changes in Mn content in the melt during growth. Sample A exhibits the highest Mn concentration (13.5 ± 1.9 at.%), which significantly decreases in Sample B (7.2 ± 1.1 at.%). This reduction is addressed in the *Discussion of Growth* section. In Sample C, the concentration further declines to 1.14 ± 0.17 at.%, followed by a slight increase in Sample D (1.9 ± 0.3 at.%) (Table 1).

The Mn concentration within SLs remains high in both early-stage samples, A and B, at approximately 14 at.%, consistent with the nominal composition of $MnBi_2Te_4$. Toward the crystal end, a pronounced decrease in Mn content is observed: from 12-7 at.% in Sample C to 4.9 ± 0.8 at.% in Sample D. As discussed in our previous work

[5], Mn depletion in SLs severely degrades both the magnetic properties and the electronic structure of surface states.

Mn may also be present in the QLs, as indicated by our earlier studies [5], potentially mediating ferromagnetic coupling between distant SLs via Mn-doped quintuple blocks. In the present study, the Mn concentration in larger QL blocks is approximately 0.7 at.%. However, between SLs, accurate quantification is challenging due to the close proximity of Mn-rich layers, which may lead to an overestimation of the measured Mn concentration.

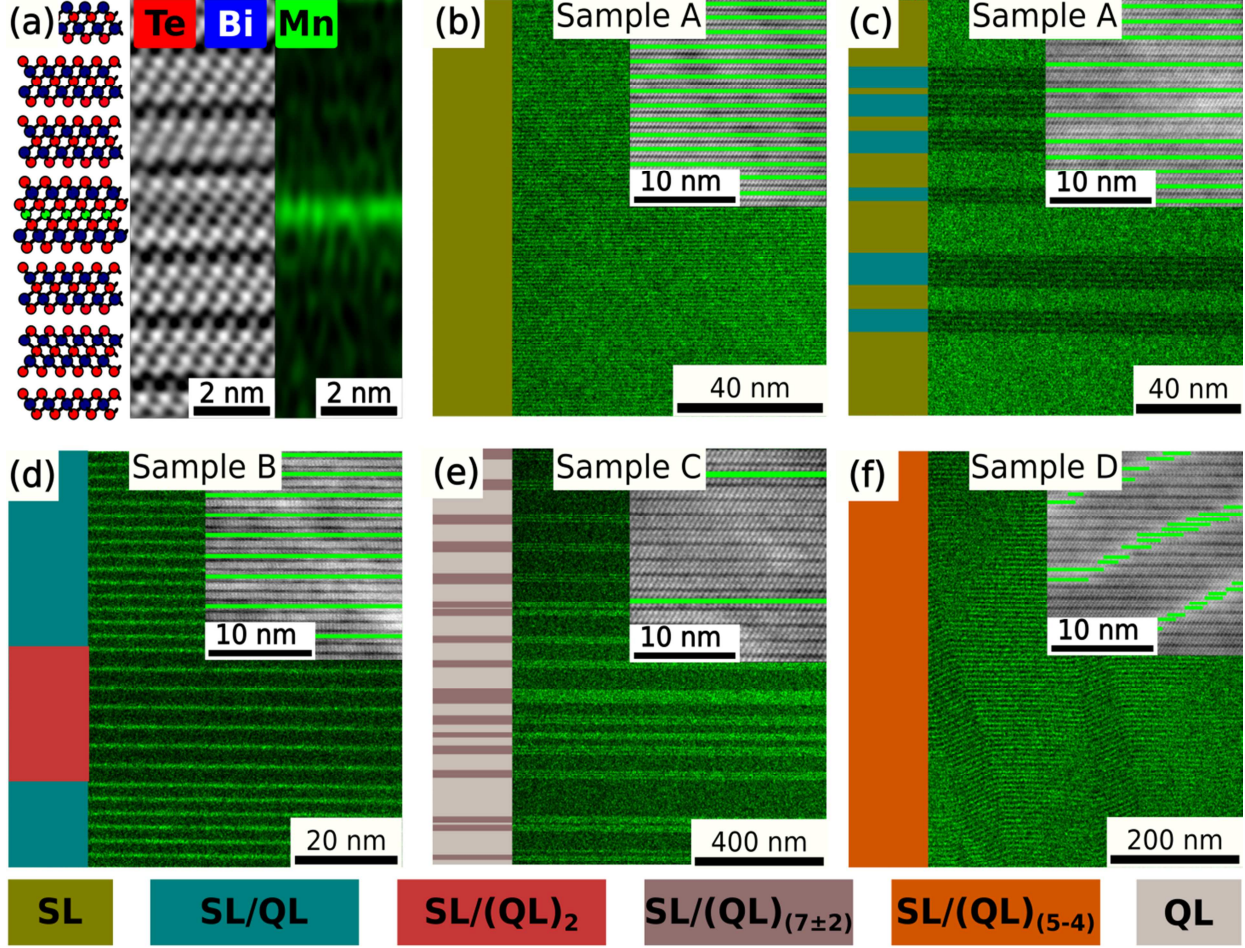

**Fig. 1.** (a) Crystal structure of $Bi_2Te_3$ quintuple layers (QLs) and $MnBi_2Te_4$ septuple layers (SLs) separated by van der Waals gaps: a schematic illustration (left), visualized using transmission electron microscopy (middle) and Energy-Dispersive X-ray mapping (right). (b)-(f) Energy-Dispersive X-ray elemental maps showing the spatial distribution of Mn (green) in Samples A-D, respectively. Insets present high-resolution Transmission Electron Microscopy images, with Mn in SLs marked in green. In Sample A, the $MnBi_2Te_4$ phase (b) coexists with the $MnBi_2Te_4/Bi_2Te_3$ (c). In Sample B, the $MnBi_2Te_4/Bi_2Te$ phase dominates, alongside a secondary $MnBi_2Te_4/(Bi_2Te_3)_2$ phase (d). Sample C exhibits a distribution of $MnBi_2Te_4$ layers within the $Bi_2Te_3$ matrix, with an average interlayer distance of (7 ± 2) QLs separated by larger $Bi_2Te_3$ blocks (e). Sample D shows regions with septuple layer spacing of 4 - 5 QLs, as well as areas where $MnBi_2Te_4$ layers intersect the $Bi_2Te_3$ matrix at an angle (f).

**Table 1.** Overview of sample characteristics. $T_C$ stays for Curie temperature, $T_N$ Néel temperature.

| Label | Distance from the bottom of the ampule (mm) | Material flow in the melt | Layer sequence | Average Mn concentration (at.%) | Mn concentration in SLs (at.%) | Critical temperature of dominant phases |
|---|---|---|---|---|---|---|
| A | 34-36 | Turbulent | $MnBi_2Te_4$ & $MnBi_2Te_4/Bi_2Te_3$ | 13.5 ± 1.9 | The same as average | $T_N$ = 26 K $T_N$ = 13.5 K |
| B | 20-22 | Stationary | $MnBi_2Te_4/Bi_2Te_3$ & $MnBi_2Te_4/(Bi_2Te_3)_2$ | 7.2 ± 1.1 | 14.9 ± 2.1 | $T_N$ = 13.5 K |
| C | 10-13 | Stationary | Groups of $MnBi_2Te_4/(Bi_2Te_3)_n$ $n = 7 ± 2$ | 1.14 ± 0.17 | 12 - 7 | $T_C$ = 11 K |
| D | 0-3 | No flow - diffusive transport | $MnBi_2Te_4/(Bi_2Te_3)_{4-5}$ | 1.9 ± 0.3 | 4.9 ± 0.8 | $T_C$ = 9 K |

**Magnetic properties studied using SQUID magnetometry and magnetic resonance techniques**

Magnetic properties of $MnBi_2Te_4/(Bi_2Te_3)_n$ depend critically on the stacking sequence of QLs and SLs. Density functional theory and Monte Carlo simulations [2] show that Mn magnetic moments in an isolated $MnBi_2Te_4$ layer align out-of-plane, with a Curie temperature of the ferromagnetic (FM) phase transition of 12 K. When adjacent $MnBi_2Te_4$ layers are separated only by van der Waals gaps, their Mn moments couple antiferromagnetically across layers, forming a 3D antiferromagnetic (AFM) phase with calculated Néel temperature of 25.4 K, in agreement with experimental data [3].

The magnetic behavior changes when $MnBi_2Te_4$ SLs are separated by one or more $Bi_2Te_3$ QLs. While the Mn magnetic moments within each SL remain ferromagnetically aligned, the interlayer exchange interaction weakens with increasing SL separation [3, 4]. Owing to the weak and oscillatory behaviour with distance of exchange coupling parameters, the strength and character of the interlayer interaction—FM or AFM—cannot be reliably predicted theoretically for $n$ larger than 0 ($n = 0$ - pure $MnBi_2Te_4$ phase) [4]. Experimental studies showed that inserting a single $Bi_2Te_3$ QL between $MnBi_2Te_4$ SLs reduced the Néel temperature to approximately 13 K, while increasing the separation to two QLs further lowered it to about 11.9 K. For larger separations ($n > 2$), the interlayer magnetic coupling became negligible [4].

Structural disorder—particularly Mn/Bi intermixing—is an important contributor to the magnetic properties of $MnBi_2Te_4$-based systems [5, 20–22]. This type of disorder can give rise to unexpected magnetic behaviour, including FM interlayer coupling of SLs across several QLs mediated by sparsely distributed Mn atoms substituting Bi in $Bi_2Te_3$. This phenomenon was demonstrated in our earlier work through ferromagnetic resonance measurements, interpreted within the framework of coupled magnetic thin films [5, 10]. Another, particularly detrimental form of disorder is the depletion of Mn within SLs, which degrades both the band structure of surface states and the critical temperature. In disordered FM samples, the highest Curie temperature of 12 K was achieved when the Mn content in SLs approached the nominal 14 at.%; a reduction to 4 at.% resulted in a significantly lower Curie temperature of 6 K [5].

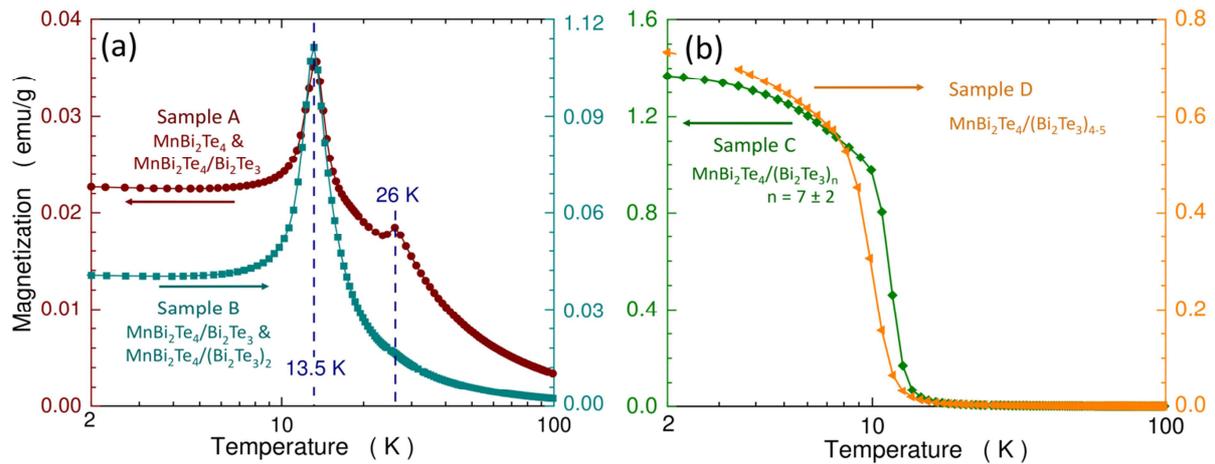

**Fig. 2.** Field-cooled magnetization of $MnBi_2Te_4/(Bi_2Te_3)_n$ measured at 100 Oe for (a) antiferromagnetic and (b) ferromagnetic samples. (a) Sample A exhibits two antiferromagnetic phases with Néel temperatures $T_N = 26$ K and $T_N = 13.5$ K attributed to $MnBi_2Te_4$ and $MnBi_2Te_4/Bi_2Te_3$, respectively. Sample B, shows a predominant antiferromagnetic transition at $T_N = 13.5$ K due to dominating $MnBi_2Te_4/Bi_2Te_3$ phase. (b) Samples C and D show ferromagnetic transitions with Curie temperatures at approximately $T_C = 10$ K. These samples exhibit increased septuple layer separation: 7 ± 2 QLs (quintuple layers) in Sample C and 4-5 QLs in Sample D.

The influence of interlayer spacing on the critical temperature is clearly manifested in the present measurements. Figure 2 shows field-cooled (FC) magnetization curves measured at 100 Oe by SQUID magnetometry. In sample A, TEM analysis revealed domains of pure $MnBi_2Te_4$ as well as regions where $MnBi_2Te_4$ layers are separated by a single $Bi_2Te_3$ layer. These structural motifs correspond to two distinct antiferromagnetic transitions observed by SQUID at 26 K and 13.5 K (Fig. 2a), respectively. Although slightly higher than previously reported values [3, 4], these transitions remain consistent with the expected behavior of $MnBi_2Te_4$ and $MnBi_2Te_4/Bi_2Te_3$ phases.

Magnetic studies of Sample B, in which TEM analysis revealed the presence of both the $MnBi_2Te_4/Bi_2Te_3$ and $MnBi_2Te_4/(Bi_2Te_3)_2$ phases [Fig. 1 (d)], shows only a single peak at 13.5 K, a characteristic feature of the $MnBi_2Te_4/Bi_2Te_3$ phase. This suggests that $MnBi_2Te_4/Bi_2Te_3$ is the dominant phase in this sample.

It is noteworthy that in both these samples (A and B) the $MnBi_2Te_4$ layers remain structurally and chemically well-ordered, with Mn concentrations close to the nominal 14 at.%, indicating no significant depletion (Table 1). This contrasts with previously reported $MnBi_2Te_4/(Bi_2Te_3)_2$ systems exhibiting partial Mn loss (down to ~8 at.%) and a transition to ferromagnetic behavior, attributed to Mn/Bi intermixing between septuple and quintuple layers [5].

Samples C and D no longer exhibit the AFM peak, which is consistent with the increased separation between SLs—averaging seven QLs in Sample C and 4–5 QLs in Sample D (Table 1). Field-cooled magnetization measurements as a function of temperature (Fig. 2b) reveal overall ferromagnetic-like behavior, with a Curie temperature near 10 K, in agreement with previous reports for systems with SL separation exceeding 3 QLs [2, 4]. The observed critical temperature is commonly associated with the intrinsic $T_c$ of an individual SL [4].

Sample C shows both a higher saturation magnetization and a slightly elevated $T_c$ (~11 K) compared to Sample D (~9 K), despite its lower overall Mn content (Table 1). This behavior is likely a consequence of the higher structural and compositional quality of SLs in Sample C, where the Mn concentration within the SLs reaches ~12 at.%. In contrast, Sample D exhibits reduced Mn incorporation in the SLs (~5 at.%), which may lead to the suppression of magnetic ordering within individual layers. The observed reduction in $T_C$ is consistent with previous findings [5].

The magnetic resonance results are consistent with the magnetometry data. X-band resonance spectra measured as a function of temperature for two orientations of the external magnetic field **H** relative to the $Bi_2Te_3$ *c*-axis (**H** ∥ **c** and **H** ⊥ **c**), are shown in Supplementary Fig. 1. The corresponding anisotropy of the resonance field ($H_{res}$) is presented in Fig. 3.

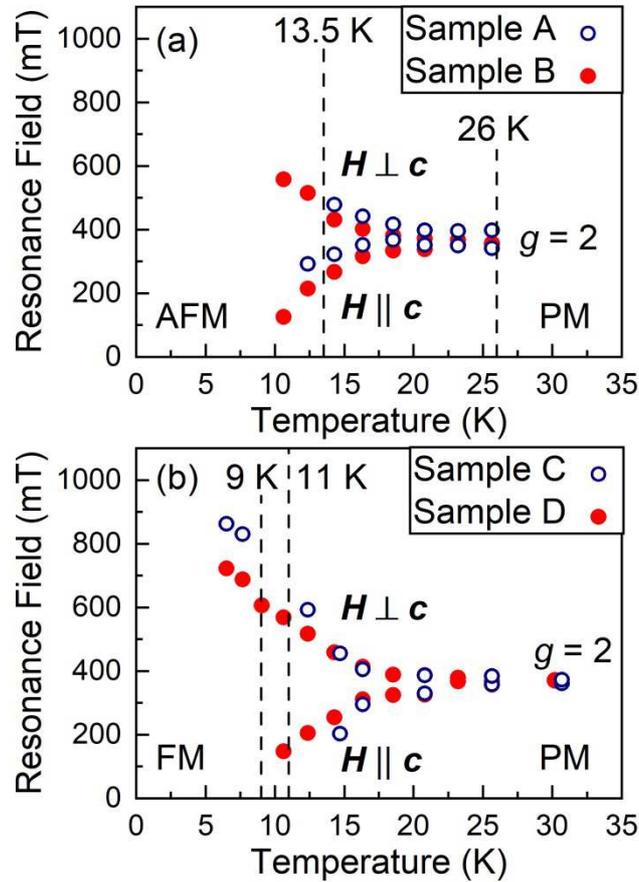

**Fig. 3.** Temperature dependence of the resonance field in $MnBi_2Te_4/(Bi_2Te_3)_n$ measured by X-band magnetic resonance for magnetic field applied parallel (**H** ∥ **c**) and perpendicular (**H** ⊥ **c**) to the $Bi_2Te_3$ *c*-axis: (a) antiferromagnetic (AFM) samples A and B; (b) ferromagnetic (FM) samples C and D. Dashed lines at 26 K and 13.5 K in (a) and at 11 K and 9 K in (b) mark the critical temperatures of the antiferromagnetic and ferromagnetic phase transitions, respectively, as determined by SQUID magnetometry. The corresponding magnetic resonance spectra are presented in Fig. 1 of the Supplementary Information. PM stands for paramagnetic phase.

In the paramagnetic phase, all samples exhibit a single Mn-related resonance line near g ≈ 2, with negligible anisotropy. Upon approaching the magnetic phase transition—antiferromagnetic for samples A and B, or ferromagnetic for samples C and D—the anisotropy significantly increases (Fig. 3).

In the FM samples C and D, ferromagnetic resonance is observed below the Curie temperature (Fig. 3b). Due to axial symmetry, the anisotropy of the resonance field follows the relation $H_{res}^{\perp} - H_{res}^{\parallel} = 3H_A$, as reported for $MnBi_2Te_4/(Bi_2Te_3)_n$ systems [5]. Here, $H_A$ denotes the magnetic anisotropy field, and the superscripts ⊥ and ∥ indicate the orientation of **H** relative to the *c*-axis. Sample C, exhibiting higher structural and chemical order, shows stronger anisotropy, with $\mu_o H_A$ ≈ 460 mT at 6 K—comparable to the highest-quality samples in [5]. In contrast, Sample D, characterized by disordered Mn distribution, exhibits a reduced anisotropy field of $\mu_0 H_A \approx$ 370 mT. Based on SQUID-derived saturation magnetization values (~1.2 emu g$^{-1}$ for C and ~0.6 emu g$^{-1}$ for D), the uniaxial anisotropy constant $K_1$ is estimated to be 4300 J m$^{-3}$ and 1700 J m$^{-3}$ for Samples C and D, respectively. The calculation procedure for $K_1$ from $H_A$ follows Ref. [5].

In AFM samples A and B, the resonance signal is observed in the paramagnetic regime but vanishes below $T_N$. Due to the complex mode structure, antiferromagnetic resonance requires multi-frequency techniques beyond the capabilities of the X-band measurements employed here. In Sample A, where magnetometry data reveal two AFM transitions at 26 K and 13.5 K, an enhanced anisotropy of the resonance line (~60 mT) is observed just above 26 K. In Sample B, this effect is weaker, reflecting the absence of a higher-$T_N$ AFM phase (Fig. 3a).

**Discussion of growth**

As discussed, the crystals comprise alternating QL and SL sequences that determine their magnetic properties, which can be tuned from antiferromagnetic to ferromagnetic by varying the interlayer spacing. This section addresses the growth mechanisms underlying the formation of these sequences and the resulting magnetic behavior. Four distinct growth stages were identified: (i) turbulent material flow, (ii) rapid MnTe precipitation, (iii) stationary flow, and (iv) cessation of material flow. These stages are discussed in detail alongside the introduction of the relevant growth concepts.

Both our observations and literature reports [23] indicate that Bi$_2$Te$_3$ and related compounds preferentially grow with their *c*-axes aligned parallel to the crystallization front. This orientation arises from the weak van der Waals interactions between external Te layers, in contrast to the chemically active edges of the QLs or SLs, which promote incorporation of molecules during growth.

Near the crystallization point, the liquid phase of bismuth telluride contains Bi$_2$Te$_3$ molecules and their derivatives [24], while MnBi$_2$Te$_4$ growth involves additional MnTe molecules. Due to similar formation enthalpies (-1.38 eV for MnTe and -0.97 eV for Bi$_2$Te$_3$ [25]), both molecules can coexist in the melt.

*Turbulent material flow.* The high melting point of MnTe ( ~ 1423 K) leads to its early precipitation during growth, depleting Mn in the melt and hindering MnBi$_2$Te$_4$ formation. This constitutes one of the main challenges in obtaining pure MnBi$_2$Te$_4$ crystals and explains why reported crystal sizes typically do not exceed a few millimetres [26, 27]. Significant MnTe precipitation observed in our earlier Horizontal Bridgman Method (HBM) experiments (unpublished) prompted us to adopt another technique - the Inverted Vertical Bridgman Method applied in this work. In this method, crystal growth begins at the ampoule's top, where *turbulent* mixing—driven by an inverted temperature gradient caused by the use of higher temperature in the lower part of the ampoule and lower temperature in the upper part—suppresses early MnTe crystallization and provides proper growth conditions for the MnBi$_2$Te$_4$ (Sample A).

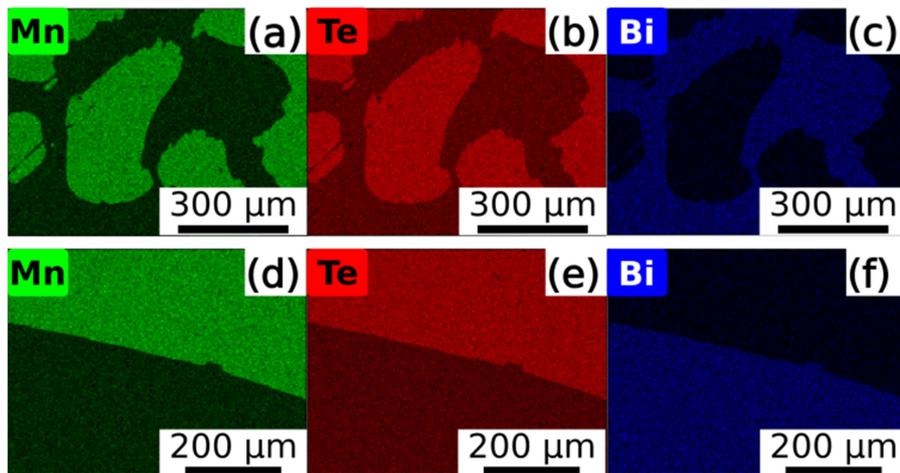

**Fig. 4.** EDX elemental maps of a crystal region formed during transition from turbulent to stationary flow in the melt. (a–c) Crystal area obtained from reduced turbulent mixing in the melt, showing Mn- and Te-rich regions indicating rapid MnTe precipitation; (d–f) The sharp interface between turbulent (top) and stationary (bottom) flow regimes, MnTe precipitation stops abruptly in the stationary flow regime.

*Rapid MnTe precipitation and initiation of stationary flow.* When the melt column height decreases to about 1.4 times its diameter, growth conditions shift abruptly from turbulent to *stationary flow,* with one or two slow vortices calmly delivering material to the surface of the growing crystal [11]. Turbulent and stationary growth regimes are separated by a transient quasi-steady state lasting several minutes. During this period, MnTe precipitates form in the growing material (Fig.4), causing a sharp decrease in Mn content—from ~ 14 at. % to ~ 7 at. % (Table 1) —both in the growing crystal and, on average, in the melt (We infer that the distribution of manganese in the melt is not uniform along the growth direction. This likely results from impurity ejection from the growing crystal, which creates a region of increased Mn concentration above its surface. We will discuss this phenomenon in more detail later, in the context of Sample C).

The MnTe precipitation results in the onset of mixed-phase MnBi$_2$Te$_4$/(Bi$_2$Te$_3$)$_n$ growth (Sample B) on a polycrystalline MnBi$_2$Te$_4$/MnTe substrate. At this stage, Mn concentration in the melt is insufficient to sustain continuous MnBi$_2$Te$_4$ formation, leading to an increased fraction of Bi$_2$Te$_3$ layers and intensified competition for Mn among the remaining MnBi$_2$Te$_4$ growth fronts.

To describe crystallization front growth of MnBi$_2$Te$_4$/(Bi$_2$Te$_3$)$_n$, we separately discuss two cases: incorporation of Bi$_2$Te$_3$ and incorporation of MnTe molecules at the layer edges.

For Bi$_2$Te$_3$ molecules, the crystallization front can be considered a *kinked surface* with numerous available bonding sites, either on Bi$_2$Te$_3$ QLs or MnBi$_2$Te$_4$ SLs (Fig. 5a). As a result, Bi$_2$Te$_3$ molecules have a similar probability of incorporating into either QLs or SLs. Surface diffusion is thus negligible, and growth is governed primarily by the melt to surface flux.

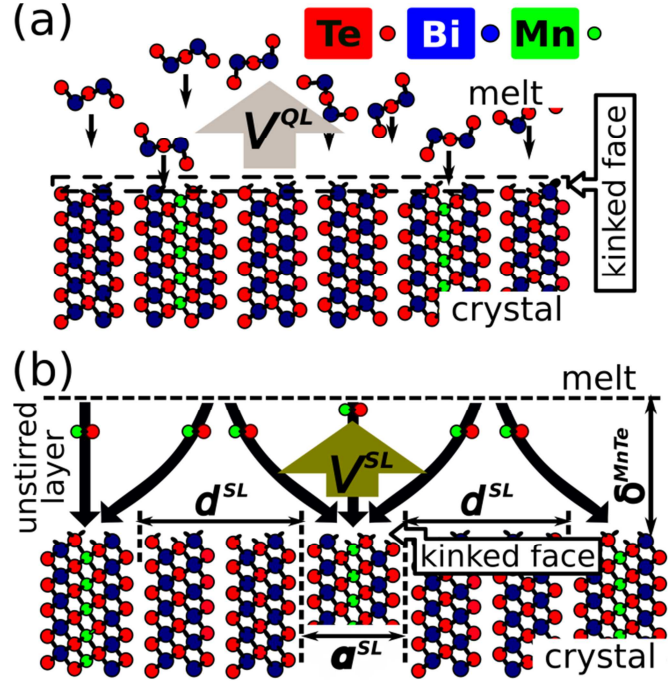

**Fig. 5**. Schematic illustration of the liquid-solid interface. (a) The flux of Bi$_2$Te$_3$ molecules in the melt determines the advance velocity $V^{QL}$ of Bi$_2$Te$_3$ layers. (b) Block diagram of Chernov's model for MnBi$_2$Te$_4$ layer growth from a solution of MnTe molecules in a Bi$_2$Te$_3$ melt. $V^{SL}$ – advance velocity of MnBi$_2$Te$_4$ layers; $\delta$ – thickness of the unstirred layer; $a^{SL}$ – width of the MnBi$_2$Te$_4$ layer; $d^{SL}$ – distance between neighboring MnBi$_2$Te$_4$ edges on the growing crystal surface.

For MnTe molecules, kinked surface growth occurs only when the Mn concentration in the melt exceeds ~14 at.%, with the crystallization front dominated by MnBi$_2$Te$_4$ edges. Upon MnTe precipitation and subsequent decrease in Mn concentration to approximately 7 at.%, the growth of the remaining MnBi$_2$Te$_4$ layers follows the solution growth mechanism consistent with Chernov's model [28], as schematically illustrated in Fig. 5b. In this regime, the growth rate-limiting factor is the *diffusion-driven transport of MnTe molecules to the MnBi$_2$Te$_4$ layer edges*. The diffusion occurs across an unstirred boundary layer of thickness $\delta$, formed above the crystal surface.

MnBi$_2$Te$_4$ edges, surrounded by Bi$_2$Te$_3$ layers that replace Mn-depleted SLs, remain active due to high density of available bonding sites, facilitating efficient MnTe incorporation. In contrast, Mn incorporation into Bi$_2$Te$_3$ is energetically unfavourable and occurs infrequently only at Bi substitutional sites or at interstitial sites [29, 30]. Consequently, isolated MnBi$_2$Te$_4$ edges act as linear, symmetric sinks of width $a^{SL}$ for MnTe molecules in the melt.

A similar case was analyzed by van der Eerden [31], who investigated the dependence of step growth velocity on the transport mechanism of growth units. By combining van der Eerden's model with Chernov's solution growth model, the following expression for the distance between the neighboring edges of MnBi$_2$Te$_4$ layers $d^{SL}$ is obtained:

$$d^{Sl} = \frac{V^{QL} a^{SL} \delta^{MnTe}}{f_0^{SL} a^{Sl} D^{MnTe} C_{eq}^{MnTe} \sigma^{MnTe} - V^{QL} \Lambda_c^{MnTe}}, \qquad (1)$$

where $V^{QL}$ is advance velocity of QL growth, $f_0^{SL}$ is the area occupied by one MnBi$_2$Te$_4$ molecule, $D^{MnTe}$ is the volume diffusion constant of MnTe molecule in the melt, $C_{eq}^{MnTe}$ is the concentration of MnTe in equilibrium with the edge of the MnBi$_2$Te$_4$ layer, $\Lambda_c^{MnTe}$ is the characteristic length for the melt-edge exchange. A more detailed description of the formula is provided in Supplementary Information.

Equation (1) describes the mechanism of *diffusion-driven competition* between MnBi$_2$Te$_4$ edges. While the amount of MnTe required for MnBi$_2$Te$_4$ layer formation remains constant, the MnTe supersaturation in the melt ($\sigma^{MnTe}$) decreases during growth due to MnTe precipitation and Mn incorporation into Bi$_2$Te$_3$. As a result, the growing MnBi$_2$Te$_4$ layer must acquire MnTe from a larger volume within the unstirred boundary layer of thickness $\delta^{MnTe}$. This leads to an *inverse relationship* between the interlayer spacing $d^{SL}$ and supersaturation $\sigma^{MnTe}$, as expressed in Eq. (1). This phenomenon is clearly observed in Sample C, where a reduced Mn concentration to 1.14 at.% leads to an increase in the distance between SLs to 7 ± 2 QLs.

In the growth model applied here, the unstirred layer thickness $\delta^{MnTe}$ plays a key role and is approximated by [32]:

$$\delta^{MnTe} = \frac{3}{2}\left(\rho^{flu} D^{MnTe}/\mu^{flu}\right)^{\frac{1}{3}}\left(\mu^{flu} X/\rho^{flu} v^{flu}\right)^{\frac{1}{2}}, \qquad (2)$$

where $\mu^{flu}$ is the viscosity of the fluid-melt, $\rho^{flu}$ the fluid-melt density, $v^{flu}$ the fluid-melt velocity, and $X$ the length of the crystal. Since $\delta^{MnTe}$ is inversely proportional to $v^{flu}$, turbulent flow in the initial stage of growth reduces the near-surface unstirred layer thickness. This represents a limiting case of the diffusion-driven competition between MnBi$_2$Te$_4$ edges, facilitating the formation of a pure MnBi$_2$Te$_4$ phase (Sample A).

*Cessation of material flow*. In the final stage of crystal growth, when the melt height drops to one-third of the crucible diameter, melt flow ceases and transport becomes *purely diffusive*. This enhances MnTe incorporation into the crystal and intensifies diffusion-driven competition between MnBi$_2$Te$_4$ layer edges (Sample D).

To explain the increased Mn concentration in Sample D compared to Sample C (Table 1), we propose the following. During the early stages of growth, the Mn concentration in the melt just above the crystallization front slightly exceeds that in the crystal, due to impurity ejection [1]. The accumulation of a Mn-enriched layer near the crystallization front results from the reluctance of Mn atoms to incorporate into Bi$_2$Te$_4$ layers. As the crystal grows, the number of Bi$_2$Te$_4$ layer edges at the front increases. This amplifies the Mn ejection effect, which intensifies toward the final stage of growth. When convection ceases, concentrations in the melt and crystal equilibrate. This leads to rapid Mn incorporation and a higher likelihood of defect formation in Bi$_2$Te$_3$ layers.

The thickness mismatch between MnBi$_2$Te$_4$ septuple layers (1.08 nm) and Bi$_2$Te$_3$ quintuple layers (0.76 nm) causes local lattice deformation when new MnBi$_2$Te$_4$ edges appear at a Bi$_2$Te$_3$-dominated front. The estimated energy required to accommodate a MnBi$_2$Te$_4$ unit at such a front ( ~ 5 eV, see Supplementary Information) far exceeds the energy gained from enthalpy of formation (~ 5 meV [25]), making spontaneous nucleation of MnBi$_2$Te$_4$ edges at an ideal Bi$_2$Te$_3$ front energetically unfavorable. However, lattice defects induced by rapid doping can serve as nucleation sites for new MnBi$_2$Te$_4$ edges. This is particularly evident in Sample D, where a high defect density promotes frequent edge skipping between MnBi$_2$Te$_4$ and Bi$_2$Te$_3$ layers, and strong diffusion-driven competition synchronizes the advancement of neighboring MnBi$_2$Te$_4$ edges, producing the characteristic structural pattern observed in Fig. 1f.

Mn doping in Bi$_2$Te$_3$ layers remains below 1 at.% throughout the crystal. This is also a natural consequence of the diffusion-driven competition between MnBi$_2$Te$_4$ edges, which preferentially incorporate MnTe from the melt, limiting its availability for Bi$_2$Te$_3$ doping. In our previous study [5], a similar Mn concentration in QLs was reported in crystals grown by the HBM, where growth occurred in a shallow trough, suppressing melt mixing. These conditions resemble the final stage of the IVBM process (no flow).

In the HBM process, a pure MnBi$_2$Te$_4$ phase was never obtained. During the initial stages of horizontal growth, most of the MnTe precipitated out, significantly lowering its concentration in the melt. As a result, in the Mn-depleted melt, diffusion-driven competition led to the formation of alternating MnBi$_2$Te$_4$ and Bi$_2$Te$_3$ layers, with an increasing number of Bi$_2$Te$_3$ layers as the MnTe content continued to decline. Furthermore, the reduced Mn availability in the melt resulted in a higher depletion of Mn within the MnBi$_2$Te$_4$ layers, as diffusion was insufficient to supply the necessary MnTe. This Mn deficiency is clearly observed both in the final stages of IVBM growth (Table 1, Sample D) and throughout the crystal grown by Horizontal Bridgman Method [5].

## Conclusions

Obtaining high-quality crystals of magnetic topological insulators via Bridgman methods faces several challenges, largely governed by the thermodynamics of growth. Nevertheless, these techniques continue to outperform epitaxial approaches by yielding materials with significantly lower free charge carrier concentrations [33].

In this study, we demonstrated that the growth of MnBi$_2$Te$_4$/(Bi$_2$Te$_3$)$_n$ superlattices via the Inverted Vertical Bridgman Method proceeds through a sequence of distinct stages, each controlled by specific thermodynamic and kinetic factors. We identified four characteristic growth regimes: (i) turbulent material flow, which enables the formation of a pure MnBi$_2$Te$_4$ phase; (ii) rapid MnTe precipitation, leading to a sharp decrease in Mn concentration in the melt (from ~14 at.% to ~7 at.%, Table 1) and initiating the incorporation of $MnBi_2Te_4$ within a $Bi_2Te_3$ matrix; (iii) a stationary flow regime that promotes the ordered assembly of QLs and SLs; and (iv) cessation of material flow, which induces structural degradation, including Mn depletion in SLs and stacking sequence inversion.

A key finding was the inverse relationship between SL spacing and MnTe supersaturation in the melt, attributed to diffusion-limited competition between septuple layers during growth. We further established a direct correlation between structural characteristics and magnetic behavior: antiferromagnetic ordering was observed in samples with closely spaced and compositionally complete SLs—namely in pure $MnBi_2Te_4$ and $MnBi_2Te_4/Bi_2Te_3$—whereas ferromagnetic signatures emerged in structures with larger SL separations (n = 4–5 and n = 7 ± 2), associated with Mn-deficient septuple layers. We determined corresponding phase transition temperatures and magnetic anisotropy constants.

Importantly, $MnBi_2Te_4/(Bi_2Te_3)_n$ crystals exhibit long-range stacking order, preserved over lateral dimensions of tens of micrometers [5]. This ordering is stabilized by the substantial elastic energy difference between SLs and QLs, which inhibits the substitution of QLs by SLs during growth without defect contribution. Moreover, the SL spacing remains constant throughout growth, provided that MnTe supersaturation is maintained in the melt. These factors underscore the self-organizing nature of this system.

These results highlight the importance of precise control over melt composition and flow dynamics in achieving structurally coherent magnetic topological insulators via bulk growth methods. The insights gained provide a foundation for tuning layer spacing, magnetic properties, and topological behavior through controlled crystal synthesis.

## References

[1] W. A. Tiller, K. A. Jackson, J. W. Rutter, B. Chalmer, *Acta Metallurgica*, 1953, **1**, 428.
[2] M. M. Otrokov, I. P. Rusinov, M.Blanco-Rey, M. Hoffmann, A. Y. Vyazovskaya, S. V. Eremeev, A. Ernst, P. M. Echenique, A. Arnau, E. V. Chulkov, *Phys. Rev. Lett.*, 2019, **122**, 107202.
[3] M. M. Otrokov, I. I. Klimovskikh, H. Bentmann, D. Estyunin, A. Zeugner, Z. S. Aliev, S. Gaß, A. U. B. Wolter, A. V. Koroleva, A. M. Shikin, M. Blanco-Rey, M. Hoffmann, I. P. Rusinov, A. Yu. Vyazovskaya, S. V. Eremeev, Yu. M. Koroteev, V. M. Kuznetsov, F. Freyse, J. Sánchez-Barriga, I. R. Amiraslanov, M. B. Babanly, N. T. Mamedov, N. A. Abdullayev, V. N. Zverev, A. Alfonsov, V. Kataev, B. Büchner, E. F. Schwier, S. Kumar, A. Kimura, L. Petaccia, G. Di Santo, R. C. Vidal, S. Schatz, K. Kißner, M. Ünzelmann, C. H. Min, Simon Moser, T. R. F. Peixoto, F. Reinert, A. Ernst, P. M. Echenique, A. Isaeva, E. V. Chulkov, *Nature*, 2019, **576**, 416.
[4] I. I. Klimovskikh, M. M. Otrokov, D. Estyunin, S. V. Eremeev, S. O. Filnov, A. Koroleva, E. Shevchenko, V. Voroshnin, A. G. Rybkin, I. P. Rusinov, M. Blanco-Rey, M. Hoffmann, Z. S. Aliev, M. B. Babanly, I. R. Amiraslanov, N. A. Abdullayev, V. N. Zverev, A. Kimura, O. E. Tereshchenko, K. A. Kokh, L. Petaccia, G. Di Santo, A. Ernst, P. M. Echenique, N. T. Mamedov, A. M. Shikin, E. V. Chulkov, *npj Quantum Mater.*, 2020, **5**, 54.
[5] J. Sitnicka, K. Park, P. Skupiński, K. Grasza, A. Reszka, K. Sobczak, J. Borysiuk, Z. Adamus, M. Tokarczyk, A. Avdonin, I. Fedorchenko, I. Abaloszewa, S. Turczyniak-Surdacka, N. Olszowska, J. Kołodziej, B. J. Kowalski, H. Deng, M. Konczykowski, L. Krusin-Elbaum, A. Wołoś, *2D Mater.*, 2022, **9**, 015026.
[6] R. S. K. Mong, A. M. Essin, J. E. Moore, *Phys. Rev. B*, 2010, **81**, 245209.
[7] N. Nagaosa, J. Sinova, S. Onoda, A. H. MacDonald, N. P. Ong, *Rev. Mod. Phys.*, 2010, **82**, 1539.
[8] M. Mogi, M. Kawamura, R. Yoshimi, A. Tsukazaki, Y. Kozuka, N. Shirakawa, K. S. Takahashi, M. Kawasaki, Y. Tokura, *Nat. Mater.*, 2017, **16**, 516.
[9] S. Grauer, K. M. Fijalkowski, S. Schreyeck, M. Winnerlein, K. Brunner, R. Thomale, C. Gould, L. W. Molenkamp, *Phys. Rev. Lett.*, 2017, **118**, 246801.
[10] H. Deng, Z. Chen, A. Wołoś, M. Konczykowski, K. Sobczak, J. Sitnicka, I. V. Fedorchenko, J. Borysiuk, T. Heider, Ł. Pluciński, K. Park, A. B. Georgescu, J. Cano, L. Krusin-Elbaum, *Nature Physics*, 2021, **17**, 36.
[11] K. Grasza, A. Jedrzejczak, *J. Crystal Growth*, 1991, **110**, 867.
[12] K. P. Kluczyk, K. Gas, M. J. Grzybowski, P. Skupiński, M. A. Borysiewicz, T. Fas, J. Suffczyński, J. Z. Domagala, K. Grasza, A. Mycielski, M. Baj, K. H. Ahn, K. Výborný, M. Sawicki, M. Gryglas-Borysiewicz, *Phys. Rev. B*, 2024, **110**, 155201.
[13] M. Sawicki, W. Stefanowicz, A. Ney, *Semicond. Sci. Technol.*, 2011, **26**, 064006.
[14] W. B. Pearson, *Canad. J. Phys.*, 1957, **35**, 886.
[15] J. R. Drabble, C. H. L. Goodman, *J. Phys. Chem. Solids*, 1958, **5**, 142.
[16] J. B. Wiese, L. Mulduwer, *J. Phys. Chem. Solids*, 1960, **15**, 13.
[17] D. S. Lee, T. H. Kim, Ch. H. Park, Ch. Y. Chung, Y. S. Lim, W. S. Seo, H. H. Park, *CrystEngComm.*, 2013, **15**, 5532.
[18] Z. S. Aliev, I. R. Amiraslanov, D. I. Nasonova, A. V. Shevelkov, N. A. Abdullayev, Z. A. Jahangirli, E. N. Orujlu, M. M. Otrokov, N. T. Mamedov, M. B. Babanly, E. V. Chulkov, *J. Alloys Compd.*, 2019, **789**, 443.
[19] Y. L. Chen, J. G. Analytis, J.-H. Chu, Z. K. Liu, S.-K. Mo, X. L. Qi, H. J. Zhang, D. H. Lu, X. Dai, Z. Fang, S. C. Zhang, I. R. Fisher, Z. Hussain, Z.-X. Shen, *Science*, 2009, **325**, 178.
[20] Y. Liu, L.-L. Wang, Q. Zheng, Z. Huang, X. Wang, M. Chi, Y. Wu, B. C. Chakoumakos, M. A. McGuire, B. C. Sales, W. Wu, J. Yan, *Phys. Rev. X*, 2021, **11**, 21033.
[21] C. Yan, Y. Zhu, L. Miao, S. Fernandez-Mulligan, E. Green, R. Mei, H. Tan, B. Yan, C.-X. Liu, N. Alem, Z. Mao, S. Yang, *Nano Letters*, 2022, **22**, 9815.

[22] X. Yao, Q. Cui, Z. Huang, X. Yuan, H. T. Yi, D. Jain, K. Kisslinger, M.-G. Han, W. Wu, H. Yang, S. Oh, *Nano Letters*, 2024, **24**, 9923.
[23] K. A. Kokh, S. V. Makarenko, V. A. Golyashov, O. A. Shegai, O. E. Tereshchenko, *CrystEngComm.*, 2014, **16**, 581.
[24] S. I. Mudry, *J. Alloys Compd.*, 1998, **267**, 100.
[25] M.-H. Du, J. Yan, V. R. Cooper, M. Eisenbach, *Adv. Funct. Mater.*, 2020, 2006516, and supplement.
[26] A. Zeugner, F. Nietschke, A. U. B. Wolter, S. Gaß, R. C. Vidal, T. R. F. Peixoto, D. Pohl, C. Damm, A. Lubk, R. Hentrich, S. K. Moser, C. Fornari, Ch. H. Min, S. Schatz, K. Kißner, M. Ünzelmann, M. Kaiser, F. Scaravaggi, B. Rellinghaus, K. Nielsch, C. Hess, B. Büchner, F. Reinert, H. Bentmann, O. Oeckler, T. Doert, M. Ruck, A. Isaeva, *Chem. Mater.,* 2019, **31**, 2795.
[27] H. Li, Sh. Liu, Ch. Liu, J. Zhang, Y. Xu, R. Yu, Y. Wu, Y. Zhang, Sh. Fan, *Phys. Chem. Chem. Phys.*, 2020, **22**, 556.
[28] A. A. Chernov, *Soviet Phys. Usp.*, 1961, **4**, 116.
[29] J. Růžička, O. Caha, V. Holý, H. Steiner, V. Volobuiev, A. Ney, G. Bauer, T. Duchoň, K. Veltruská, I. Khalakhan, V. Matolín, E. F. Schwier, H. Iwasawa, K. Shimada, G. Springholz, *New J. Phys.*, 2015, **17**, 013028.
[30] A. Wolos, A. Drabinska, J. Borysiuk, K. Sobczak, M. Kaminska, A. Hruban, S. G. Strzelecka, A. Materna, M. Piersa, M. Romaniec, R. Diduszko, *J. Magn. Magn. Mater.*, 2016, **419**, 301.
[31] J. P. van der Eerden, *J. Crys. Growth*, 1982, **56**, 174.
[32] G. H. Gilmer, R. Ghez, N. Cabrera, *J. Crys. Growth*, 1971, **8**, 79.
[33] L. Zhao, M. Konczykowski, H. Deng, I. Korzhovska, M. Begliarbekov, Z. Chen, E. Papalazarou, M. Marsi, L. Perfetti, A. Hruban, A. Wołoś, L. Krusin-Elbaum, *Nature Comm.*, 2016, **7**, 10957.

# Self-organization mechanism in Bridgman-grown MnBi$_2$Te$_4$/(Bi$_2$Te$_3$)$_n$: influence on layer sequence and magnetic properties

Paweł Skupiński,*[a] Kamil Sobczak,[b] Katarzyna Gas,[a,d] Anna Reszka,[a] Yadhu K. Edathumkandy,[a] Jakub Majewski,[c] Krzysztof Grasza,[a] Maciej Sawicki,[a,e] and Agnieszka Wołoś.[c]

a. Institute of Physics, Polish Academy of Sciences, Aleja Lotników 32/46, PL-02668 Warsaw, Poland.
b. Faculty of Chemistry, Biological and Chemical Research Centre, University of Warsaw, ul. Żwirki i Wigury 101, 02-089 Warsaw, Poland
c. Faculty of Physics, University of Warsaw, Pasteura 5, 02-093 Warsaw, Poland.
d. Center for Science and Innovation in Spintronics, Tohoku University, 2-1-1 Katahira, Aoba-ku, Sendai 980-8577, Japan.
e. Laboratory for Nanoelectronics and Spintronics, Research Institute of Electrical Communication, Tohoku University, 2-1-1 Katahira, Aoba-ku, Sendai 980-8577, Japan

* Author to whom any correspondence should be addressed.
E-mail: pawel.skupinski@ifpan.edu.pl

## Magnetic resonance

Figure 1 shows raw magnetic resonance spectra recorded in the X-band: antiferromagnetic resonance for Samples A and B, and ferromagnetic resonance for samples C and D, respectively.

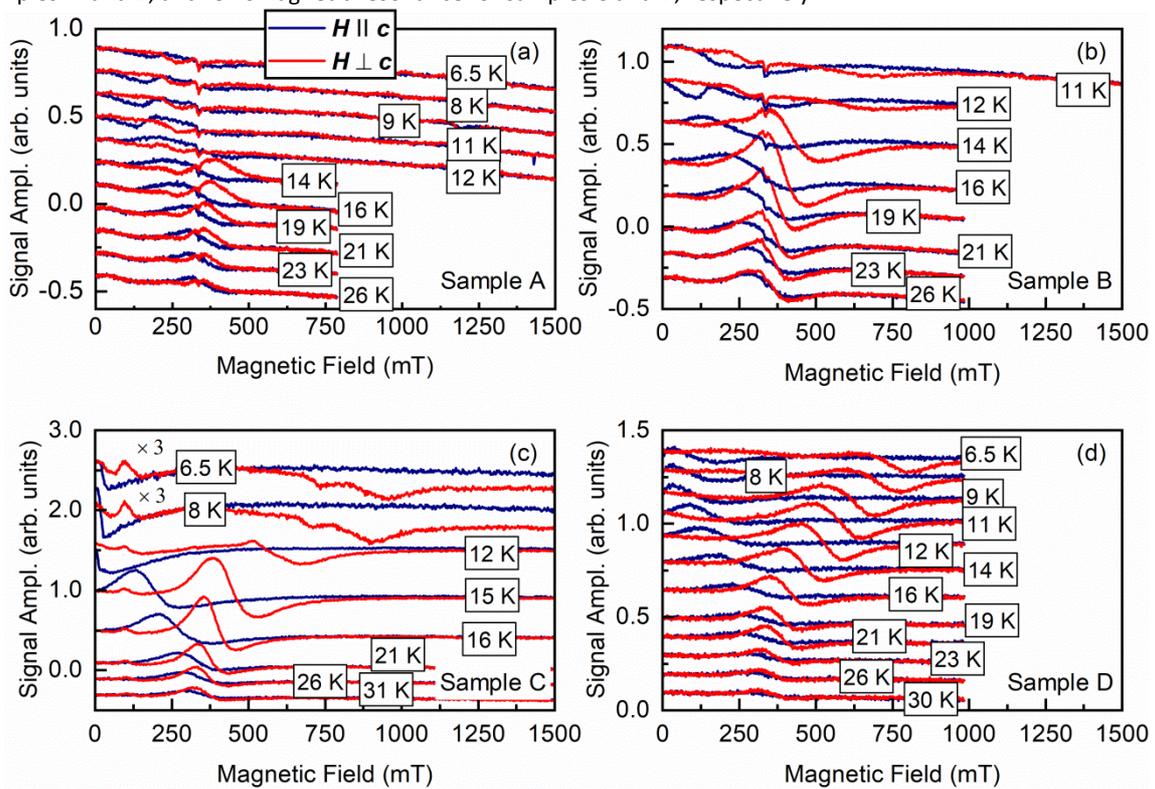

**Fig. 1.** (a)-(d) X-band magnetic resonance spectra of samples A–D, respectively, as a function of temperature, with the magnetic field applied parallel (blue lines) and perpendicular (red lines) to the Bi$_2$Te$_3$ *c*-axis. The temperature corresponding to each spectrum is indicated in boxes.

## Derivation of the expression for the distance between adjacent edges of MnBi$_2$Te$_4$ layers at the crystallization front

Variations in the crystal structure along the ingot grown by the Inverted Vertical Bridgeman Method are associated with changes in the flow dynamics of the melt from which the crystal is formed. Turbulent flow promotes the pure MnBi$_2$Te$_4$ phase, whereas laminar flow and pure diffusion transport results in the formation of mixed (MnBi$_2$Te$_4$)/(Bi$_2$Te$_3$)$_n$ structures. The A. A. Chernov model for solution growth [1] can be applied here to explain septuple layer (SL) growth in Bi$_2$Te$_3$ matrix, which is governed by transport of growth units in the liquid phase. For clarity, we restrict the discussion to the MnTe transport to the edges of MnBi$_2$Te$_4$, as this process represents the growth rate limiting factor for the formation of such layers. The edges of MnBi$_2$Te$_4$ constitute a series of parallel linear sinks on the crystal surface (see Fig. 2). Above the surface, an unstirred boundary layer of thickness $\delta^{MnTe}$ develops, across which diffusion occurs. At the boundary with the melt, the following condition applies:

$$C^{MnTe} = C_{eq}^{MnTe}(1 + \sigma^{MnTe}), \qquad (1)$$

where $C^{MnTe}$ is the concentration of MnTe molecules in the melt at a distance $\delta^{MnTe}$ from the crystal surface, $C_{eq}^{MnTe}$ is the concentration of MnTe in equilibrium with the edge of the MnBi$_2$Te$_4$ layer, and $\sigma^{MnTe}$ is the relative supersaturation that is, how many times the concentration $C^{MnTe}$ exceeds the equilibrium concentration $C_{eq}^{MnTe}$.

For the case described by the Chernov model, van der Eerden [2] derived the following expression for the advance velocity of parallel steps, $V^{SL}$, which in the present context corresponds to the advance velocity of MnBi$_2$Te$_4$ layers:

$$V^{SL} \approx \frac{\frac{f_0^{SL} a^{SL} D^{MnTe} C_{eq}^{MnTe} \sigma^{MnTe}}{\Lambda_c^{MnTe}}}{1 + \frac{a^{SL}}{\pi \Lambda_c^{MnTe}}\left[\ln\left(\frac{d^{SL}}{a^{SL}}\right) + \frac{\pi \delta^{MnTe}}{d^{SL}}\right]}. \qquad (2)$$

Here, $f_0^{SL}$ is the area occupied by a single MnBi$_2$Te$_4$ molecule, $a^{SL}$ is the width of a septuple layer (see Fig. 2), $D^{MnTe}$ is the volume diffusion constant of an MnTe molecule in the melt, $\Lambda_c^{MnTe}$ is the characteristic length for the melt-edge exchange, $d^{SL}$ is the spacing between adjacent MnBi$_2$Te$_4$ edges.

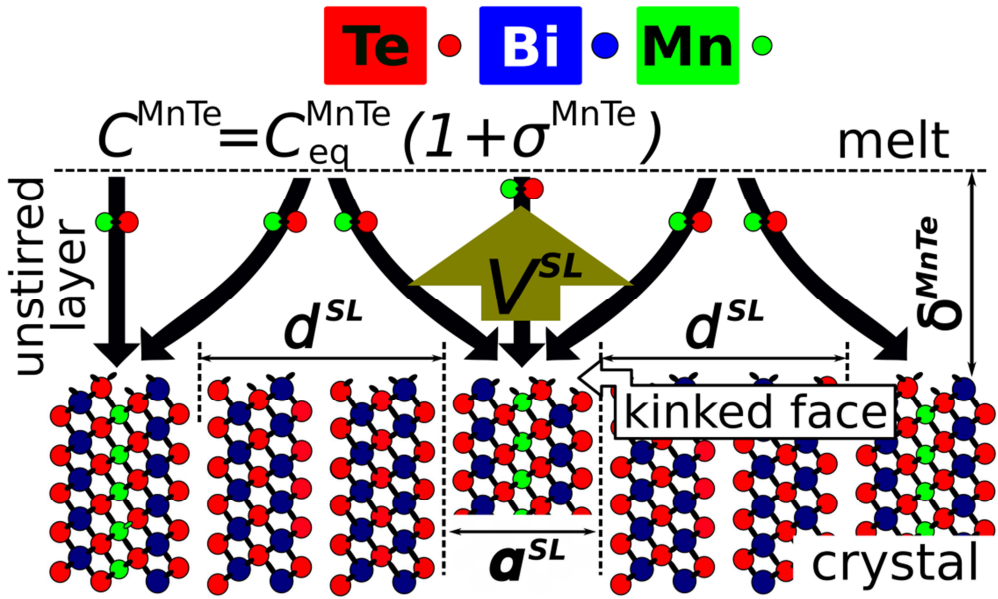

**Fig. 2.** Schematic illustration of the MnTe particle flow toward the edges of MnBi$_2$Te$_4$ layers on the surface of the growing crystal. $V^{SL}$ – advance velocity of MnBi$_2$Te$_4$ layers; $\delta^{MnTe}$ – thickness of the unstirred layer; $C^{MnTe}$ – concentration of MnTe molecules in the melt at a distance $\delta^{MnTe}$; $C_{eq}^{MnTe}$ – concentration of MnTe in equilibrium with the edge of the MnBi$_2$Te$_4$ layer, $\sigma^{MnTe}$ – relative supersaturation at a distance $\delta^{MnTe}$, $a^{SL}$ – width of a MnBi$_2$Te$_4$ layer, $d^{SL}$ – spacing between neighboring MnBi$_2$Te$_4$ edges on the surface of the growing crystal.

In both Bi$_2$Te$_3$ and MnBi$_2$Te$_4$ the quintuple and septuple layers tend to grow perpendicular to the crystallization front, as demonstrated by experiments conducted in our laboratory and elswhere [3]. For the MnBi$_2$Te$_4$/(Bi$_2$Te$_3$)$_n$ structure to form, the advance velocities of quintuple ($V^{QL}$) and septuple ($V^{SL}$) layers must be equal:

$$V^{SL} = V^{QL}. \qquad (3)$$

We now substitute $V^{SL}$ for expression (2), and, for simplicity, neglect the factor $\ln(d^{SL}/a^{SL})$. Preliminary order-of-magnitude estimates of the individual terms [1] indicate that this approximation introduces an error below 1% within the observed range of $d^{SL}$ (1.34 nm for one QL separation and 9.5 nm for eight QLs).

$$\frac{\frac{\Lambda_c^{MnTe}}{a^{SL}\delta^{MnTe}}}{\Lambda_c^{MnTe}d^{SL}} = V^{QL} \qquad (4)$$

Let us now rearrange equation (4) to derive an expression for the distance between neighboring edges of the MnBi$_2$Te$_4$ layers, $d^{SL}$:

$$d^{SL} = \frac{V^{QL} a^{SL} \delta^{MnTe}}{f_0^{SL} a^{SL} D^{MnTe} C_{eq}^{MnTe} \sigma^{MnTe} - V^{QL} \Lambda_c^{MnTe}} \qquad (5)$$

# Calculation of the energy required for the formation of a MnBi$_2$Te$_4$ layer in the Bi$_2$Te$_4$ structure

Let us estimate the energy required to create space for a single MnBi$_2$Te$_4$ molecule at the crystallization front containing edges of Bi$_2$Te$_3$. This energy corresponds to work needed to separate two Bi$_2$Te$_3$ quintuple layers against the elastic force. For simplicity, we assume a symmetric configuration and consider the work $W$ required to deform a single quintuple layer:

$$W = \frac{1}{2}\Delta r \cdot F, \qquad (6)$$

where: $\Delta r$ is the difference in thickness between the MnBi$_2$Te$_4$ and the Bi$_2$Te$_3$ layers equal to 0.32 nm and $F$ is the elastic force calculated from:

$$F = \int_0^\pi F_f \, d\varphi. \qquad (7)$$

Here, $F_f$ is the force acting on a deformed fragment with an area of longitudinal section of Bi$_2$Te$_3$ molecule, A =3.34 nm$^2$, and a deformation length $l$ (based on our observations and those reported in literature, $l \approx 1$ nm). The force $F_f$ is calculated using Hooke's law for shear stress:

$$\tau = \frac{F_f}{A} = \frac{\gamma E}{2(1+\nu)}, \qquad (8)$$

where $E$ is the experimentally determined Young's modulus for Bi$_2$Te$_3$ crystals along the cleavage plane ($E$ = 38 GP [4]), $\nu$ is Poisson's ratio for Bi$_2$Te$_3$ crystals ($\nu$ = 0.26 [5]) and $\gamma$ is the shear strain, which for small deformations can be approximated as $\gamma \approx 1/2(\Delta r/l)$.

The energy required to create a space for a single MnBi$_2$Te$_4$ molecule at the crystallization front is given by:

$$Q_{def} = 2W \approx 5 eV, \qquad (9)$$

which is a greater then the energy gain associated with the enthalpy of formation of a single molecule, approximately -5 meV [6].

## References

[1] A. A. Chernov, *Soviet Phys. Usp.*, 1961, **4**, 116.
[2] J. P. van der Eerden, *J. Crys. Growth*, 1982, **56**, 174.
[3] K. A. Kokh, S. V. Makarenko, V. A. Golyashov, O. A. Shegai, O. E. Tereshchenko, *CrystEngComm.*, 2014, **16**, 581.
[4] C. Lamuta, D. Campi, A. Cupolillo, Z.S. Aliev, M.B. Babanly, E.V. Chulkov, A. Politano, L. Pagnotta, Scr. Mater. 2016, **121**, 50.
[5] D. L. Medlin, K. J. Erickson, S. J. Limmer, W. G. Yelton, M. P. Siegal, J. Mater. Sci. 2014, **49**, 3970.
[6] M.-H. Du, J. Yan, V. R. Cooper, M. Eisenbach, *Adv. Funct. Mater.*, 2020, 2006516, and supplement.